\documentclass[12pt,draftclsnofoot,onecolumn]{IEEEtran}

\usepackage{blindtext}
\usepackage{amssymb}
\usepackage{amsmath}
\usepackage{graphicx}
\usepackage{cite}
\usepackage{citesort}
\usepackage{multicol}
\usepackage{amsfonts}
\usepackage{color}
\usepackage{subfigure}
\usepackage{epstopdf}
\usepackage{amssymb}
\usepackage{comment}
\usepackage{amsmath}
\allowdisplaybreaks[4]
\usepackage{booktabs}
\usepackage{multirow}
\usepackage{bmpsize}
\usepackage[ruled,linesnumbered]{algorithm2e}

\usepackage{lipsum}
\usepackage{stfloats}
\newtheorem{theorem}{\textbf{Theorem}}

\newtheorem{lemma}{\textbf{Lemma}}

\newtheorem{corollary}{\textbf{Corollary}}

\makeatletter
\def\ScaleIfNeeded{%
\ifdim\Gin@nat@width>\linewidth \linewidth \else \Gin@nat@width
\fi } \makeatother

\begin{document}
%\pagestyle{fancyplain}
%
%\pagestyle{fancy}
%\lhead[]%
 %   {\footnotesize Physical layer security}
%\cfoot{}

\title{\huge{Scalable Multiuser Immersive Communications with Multi-numerology and Mini-slot}}
%\author{
%\IEEEauthorblockN{Ming Hu\IEEEauthorrefmark{1}, Jiazhi Peng\IEEEauthorrefmark{1}, Yueyang Zhang\IEEEauthorrefmark{1},  Lifeng Wang\IEEEauthorrefmark{1}}
%\IEEEauthorblockA{\IEEEauthorrefmark{1} School of Information Science and Engineering, Fudan University, Shanghai, China
\author{Ming Hu, Jiazhi Peng, Lifeng Wang, and Kai-Kit Wong
%\author{Author 1, Author 2, Author 3,  Author 4, and Author 5   and Shi Jin,~\IEEEmembership{Senior Member,~IEEE}
\thanks{M. Hu, J. Peng, and L. Wang are with the School of Information Science and Engineering, Fudan University, Shanghai 200433, China (e-mail: $\rm {20210720045,21210720209,lifengwang}@fudan.edu.cn$).}
\thanks{K.-K. Wong is with the Department of Electronic and Electrical Engineering, University College London, London WC1E 7JE, U.K. (e-mail:
kai-kit.wong@ucl.ac.uk).}
}

%\IEEEauthorrefmark{2}Faculty of Information Technology and Communication Sciences, Tampere University, Finland} nanchi

\maketitle

\begin{abstract}
This paper studies multiuser immersive communications networks in which different user equipment may demand various extended reality (XR) services. In such heterogeneous networks, time-frequency resource allocation needs to be more adaptive since XR services are usually multi-modal and latency-sensitive. To this end, we develop a scalable time-frequency resource allocation method based on multi-numerology and mini-slot. To appropriately determining the discrete parameters of multi-numerology and mini-slot for multiuser immersive communications, the proposed method first presents a novel flexible time-frequency resource block configuration, then it leverages the deep reinforcement learning to maximize the total quality-of-experience (QoE) under different users' QoE constraints. The results confirm the efficiency and scalability of the proposed time-frequency resource allocation method.
\end{abstract}

\begin{IEEEkeywords}
Immersive communications, quality of experience, numerology, mini-slot.
\end{IEEEkeywords}

%========================================================================
\section{Introduction}
The sixth-generation wireless (6G) communications systems  seek to boost scalable and resilient transmissions for addressing new challenges, which come from the emerging multi-modal extended reality (XR) services. As one of 6G key use cases, immersive communications provide remote XR services including mixed reality and augmented reality, however, these immmersive services impose stringent requirements on data rate and communications latency~\cite{John_2012,huming2023}. Multi-numerology and mini-slot are compelling access technologies for supporting diverse scenarios and requirements, which have been applied in 5G systems~\cite{Zaidi2018}.

Since sub-6 GHz, millimeter wave (mmWave) and terahertz (THz) frequency bands shall be widely-used in 6G, {the frequency bandwidths at different carrier frequencies are dramatic, which may substantially increase system complexity and peak-to-average power ratio (PAPR) if the number of subcarriers  increases with adding more frequency bandwidths. In light of hardware constraints and various XR service requirements, multi-numerology is introduced to make flexible resource allocation, compared to the uniformly distributing frequency resource allocation in 5G systems~\cite{Peng2017}. Its gist is that subcarrier spacing can be tuned according to the specific frequency bandwidths and service requirements, in this way, the number of subcarriers can keep the same for different amounts of frequency bandwidths  without largely increasing the PAPR.} To meet the requirements of mission-critical services, \cite{Lei2018} proposes a scalable resource block based on flexible numerology. In~\cite{xiaopei_2018}, mixed-numerology is adopted in windowed orthogonal frequency division multiplexing (OFDM) systems. Like traditional single-numerology case, multi-numerology OFDM systems are still susceptible to the critical PAPR, hence~\cite{Memisoglu2021} proposes a numerology scheduling method to reduce the PAPR. The numerology scheduler of~\cite{Boutiba2022} is designed to be adaptive for satisfying different slices' requirements. In vehicle-to-everything networks,  recent work~\cite{Nguyen2022} shows that the use of 5G numerology can improve Quality-of-Service. On the other hand, mini-slot is an alternative for further enhancing the network's scalability, which achieves much lower latency since it allows shorter transmission slot duration to contain only a few OFDM symbols, therefore, mini-slot plays a promising role in ultra-reliable low-latency communication (URLLC). When URLLC and the enhanced mobile broadband (eMBB) traffics coexist in multiuser communications systems,~\cite{Hao2021} provides a media access control (MAC) layer scheduling for maximizing eMBB utility under URLLC latency constraint. To flexibly manage the eMBB and URLLC network slices,~\cite{Setayesh2022} investigates the numerology and mini-slot for maximizing the network throughput while satisfying service level agreement. Meanwhile, ~\cite{Schober2022} develops a hybrid puncturing and superposition scheme which can maximize the minimum average throughput of eMBB and URLLC users.

In immersive communications systems, 360{$^\circ$} video is an important component for XR applications~\cite{Yaqoob2020,huming2023}. However, 360{$^\circ$} video streaming demands large frequency bandwidths for low-latency transmissions. Existing research contributions have attempted to improve the efficiency of 360{$^\circ$} video streaming under limited resources constraints. In~\cite{Yaowang2019},  a two-tier 360-degree video streaming scheme is investigated to address the network dynamics, such a scheme is source representation independent and outperforms the DASH streaming and Field of View (FoV) streaming schemes. Since videos in the FoV region must be delivered, ~\cite{Nguyen2018} analyzes the future FoV region in a closed-form under a given confidence level. By considering the  tile-based 360-degree video streaming, ~\cite{Kaigui2019} adopts a machine learning method to optimize different Quality of Experience (QoE) objectives. A viewport-aware adaptive tiling scheme is studied in~\cite{Hongkai2022}, where the long-term user QoE is maximized. In~\cite{Chengjun2021}, energy efficiency of delivering multi-quality tiled 360 virtual reality (VR) videos is enhanced by jointly determining appropriate beamforming, subcarrier and the choices of quality level. Latest work~\cite{huming2023} designs flexible frame structures for achieving the two-tier 360-degree video delivery and improving the system robustness.

Motivated by the aforementioned studies, {this paper aims to establish the scalable two-tier 360-degree video streaming in multiuser immersive communications systems by leveraging multi-numerology and mini-slot according to the 3GPP~\cite{Zaidi2018,Ljiljana2020,TS38_211}.  To the best of our knowledge, this is the first work to exploit the merits of joint multi-numerology and mini-slot design for 360-degree video delivery. We orchestrate a flexible time-frequency resource block structure and propose a novel time-frequency allocation method to maximize the total QoE under diverse levels of UE's QoE constraint, which significantly outperforms the QoE-based greedy and equal resource allocation solutions. }

\section{System Descriptions}\label{System_description}
\begin{figure}[t!]
\centering
\includegraphics[width=3.5 in]{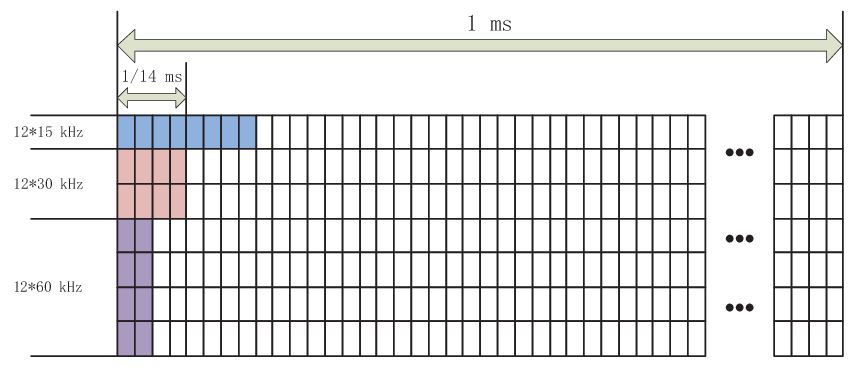}
\caption{An illustration of time-frequency resource structure with multi-numerology ($\mu \in \{0,1,2\}$ in this figure) and the same mini-slot value $\eta=2$ for all the UEs,
in which each rectangle of the resource grid represents a resource block given by \eqref{resource_block}, and each UE's allocated bandwidth part has the same number of resource blocks given by ${\rm C}^{\rm BWP}=8{\rm C}^{\rm RB}$ according to \eqref{BWP},
regardless of each UE's numerology.
}% \textcolor[rgb]{1.00,0.00,0.00}{edge clouds  at or linked to the SBSs through optical fiber}
\label{RB_numerology}
\end{figure}

In a downlink immersive communications system, base station (BS) delivers 360-degree video contents to $N$ user equipment (UEs). To meet different UEs' QoE, we manage time-frequency resources via numerology and mini-slot. According to the 5G new radio (5G NR)~\cite{Zaidi2018,Ljiljana2020,TS38_211}, each subframe duration is fixed value of 1ms, the OFDM's subcarrier spacing is $15\times2^{\mu}$kHz with the numerology $\mu$ (a non-negative integer value for low-latency XR services), and each resource block has 12 consecutive subcarriers with the same numerology in the frequency domain. Note that each slot  contains 14 OFDM symbols,  and 4G LTE's numerology is $\mu=0$ with the slot duration of 1ms~\cite{Zaidi2018}. For generality, we consider that the $n$-th UE's mini-slot duration can be selected to contain $\eta_n$ OFDM symbols, and each  slot is divided into several mini-slots whose lengths may be different. As such, we design a new time-frequency resource block configuration (the unit is ms$\times$kHz=s$\times$Hz) to achieve scalable multiuser immersive communications, which is defined as follows:
\begin{align}\label{resource_block}
{\rm C}^{\rm RB}&=\Delta T_{\rm min} \times \Delta B_{\rm min} \nonumber\\
&=\frac{1}{14\times 2^{\mu_{\rm max}}} \times \left(12 \times 15\times2^{\mu_{\rm min}}\right),
\end{align}
where $\Delta T_{\rm min}=\frac{1}{14\times 2^{\mu_{\rm max}}}$ms is the minimum OFDM symbol duration that can be supported in the system with the maximum numerology value $\mu_{\rm max}$, considering the fact that the OFDM symbol duration without cyclic prefix (CP) changes inversely to its subcarrier spacing; $\Delta B_{\rm min}=12 \times 15\times2^{\mu_{\rm min}}$kHz is the minimum bandwidth part in the frequency domain with the minimum numerology value $\mu_{\rm min}$, here, the bandwidth part is a subset of resource blocks for a given numerology~\cite{TS38_211}. The advantage of such a resource block design is that it solely depends on the system's maximum and minimum numerology values regardless of UEs' specific XR service requirements, in this case, the size of a bandwidth part (the unit is ms$\times$kHz=s$\times$Hz) occupied by the $n$-th UE with the required numerology $\mu_n$ and mini-slot $\eta_n$ can be easily calculated as
\begin{align}\label{BWP}
{\rm C}^{\rm BWP}&=\left(\eta_n\times 2^{\mu_{\rm max}-\mu_n} \Delta T_{\rm min}\right) \times \left(2^{\mu_n-\mu_{\rm min}} \Delta B_{\rm min}\right) \nonumber\\
&=\eta_n \times  2^{\mu_{\rm max}-\mu_{\rm min}}\times {\rm C}^{\rm RB}.
\end{align}
Each bandwidth part is dedicated to one UE and there is no overlap between them. From \eqref{BWP}, the number of resource blocks in each UE's allocated bandwidth part only depends on its mini-slot for a given immersive communications system, as illustrated in Fig.~\ref{RB_numerology}.

 In the two-tier 360-degree video streaming system~\cite{huming2023}, each video frame has the duration of $T$ (ms) and frequency bandwidth $B$ (kHz), which is partitioned into basic tier (BT) video chunk phase and and enhancement tier (ET) video chunk phase~\cite{Yaowang2019}. Let $\mu_n^{\rm BT}$, $\eta_n^{\rm BT}$, $\mu_n^{\rm ET}$, and $\eta_n^{\rm ET}$ denote the BT's and ET's numerology and mini-slot, respectively. Based on \eqref{BWP}, we have the following time resource constraint
\begin{align}\label{T_constr}
&\sum\limits_{(i,i')\in\mathcal{ \{L,R\}}, \hfill \atop(i,i')\notin\mathcal{ \{B, A\}}} \eta_{n,i}^{\rm BT} 2^{\mu_{\rm max}-\mu_{n,i}^{\rm BT}} \Delta T_{\rm min}+ \nonumber\\
&\quad\sum\limits_{(j,j')\in\mathcal{ \{L,R\}}, \hfill \atop(j,j')\notin\mathcal{ \{B, A\}} } \eta_{n,j}^{\rm ET} 2^{\mu_{\rm max}-\mu_{n,j}^{\rm ET}} \Delta T_{\rm min} \leq T,\;\;\; \forall n,
\end{align}
and frequency resource constraint
 \begin{align}\label{B_constr}
\hspace{-0.2 cm}\sum\limits_{(i,i')\in\mathcal{ \{B, A\}}, \hfill \atop (i,i')\notin\mathcal{ \{L,R\} }} 2^{\mu_{n,i}^\nu-\mu_{\rm min}} \Delta B_{\rm min} \leq B, \;\;\; \forall n, \nu \in\{\rm BT, \rm ET\},
\end{align}
where { $\mu_{n,i}^{\rm BT}$ and $\eta_{n,i}^{\rm BT}$ are the numerology and mini-slot of the $i$-th bandwidth part at BT for the UE $n$, respectively, $\mu_{n,j}^{\rm ET}$ and $\eta_{n,j}^{\rm ET}$   are the numerology and mini-slot of the $j$-th bandwidth part at ET for the UE $n$, respectively}, the relative positioning constraint $(i,i')\in\mathcal{ \{L,R\}}$ means that the bandwidth part $i$ is the left or right of the bandwidth part $i'$, i.e., there is no overlap between them in the time domain, the relative positioning constraint $(i,i')\in\mathcal{ \{B, A\}}$ means that the bandwidth part $i$ is the below or above of bandwidth part $i'$, i.e., there is no overlap between them in the frequency domain,

 Since the time-frequency resources are orthogonally allocated to the UEs (no co-channel interference), the BT's transmission rate (bits per frame) is given by
 \begin{align}\label{BT_rate}
R_n^{\rm BT}=\sum\limits_i {\rm{C}}_{{\rm BT},n,i}^{\rm{BWP}} \log_2\left(1+\frac{G_n^{\rm t} G_n^{\rm r} \left|\hbar_n^{\rm BT}\right|^2  p_n}{\delta^2}\right),
\end{align}
where ${\rm{C}}_{{\rm BT},n,i}^{\rm{BWP}}$ is the $i$-th bandwidth part allocated to the $n$-th UE at BT with the mini-slot $\eta_{n,i}^{\rm BT}$, $G_n^{\rm t}$ and $G_n^{\rm r}$ are the effective transmit antenna gain and receive antenna gain obtained by the UE $n$, respectively, $\left|\hbar_n^{\rm BT}\right|^2$ and $\left|\hbar_n^{\rm ET}\right|^2$ are the large-scale fading channel power gains\footnote{In the highly directional beamforming transmissions such as millimeter wave (mmWave), the effect of small-scale fading on the channel power gain could be ignored.} $p_n$ is the transmit power spectral density (PSD) of the downlink channel between the BS and the $n$-th UE,  and $\delta^2$ is the noise's PSD.
Likewise, ET's transmission rate (bits per frame) is given by
{{\begin{align}\label{ET_rate}
R_n^{\rm ET}=\sum\limits_j {\rm{C}}_{{\rm ET},n,j}^{\rm{BWP}} \log_2\left(1+\frac{G_n^{\rm t} G_n^{\rm r} \left|\hbar_n^{\rm BT}\right|^2  p_n}{\delta^2}\right),
\end{align} }}
where ${\rm{C}}_{{\rm ET},n,j}^{\rm{BWP}}$ is the  $j$-th bandwidth part allocated to the $n$-th UE at ET with the mini-slot $\eta_{n,j}^{\rm ET}$. We further highlight that the proposed resource block and bandwidth part designs facilitate the analysis of transmission rate in multiuser communications since \eqref{BT_rate} and \eqref{ET_rate} are available for arbitrary UEs' numerologies.

One of the key performance indicators (KPIs) in the two-tier video delivery system is the QoE, which is commonly-measured in a logarithmic manner~\cite{Yaowang2019,Hongkai2022}, i.e., the QoE for the $n$-th UE is calculated as
\begin{align}\label{eq2}
\widetilde{Q}_n=\left(1-\rho_n\right)Q_n\left(\hat{R}_n^{\rm BT}\right)+\rho_n Q_n\left(\Xi_n\right),
\end{align}
where $\rho_n$ ($0\leq \rho_n \leq 1$) is the probability of precisely predicting UE's FoV, $Q_n\left(x\right)=a_n+b_n\log\left(x\right)$ with specific XR video-dependent constant parameters $a_n$ and $b_n$, $\hat{R}_n^{\rm BT}=\frac{R_n^{\rm BT}}{\mathcal{C}^{\rm BT}}$ with the coverage area of the 360$^\circ$ video $\mathcal{C}^{\rm BT}$, $\Xi_n=\hat{R}_n^{\rm BT}+\hat{R}_n^{\rm ET}$ denotes the effective rate after using layered coding to generate the ET chunks, here $\hat{R}_n^{\rm ET}=\frac{R_n^{\rm ET}}{\mathcal{C}^{\rm ET}}$ with the coverage area of the ET chunk $\mathcal{C}^{\rm ET}$.

By appropriately selecting numerology and mini-slot, our aim is to maximize the total QoE while achieving the minimum QoE per active UE, thus the time-frequency resource allocation problem is formulated as
\begin{align}\label{P1}
&\mathop {\max }\limits_{\mathbf{x},\boldsymbol{\mu,\eta}}  \sum\limits_{n = 1}^N  x_n \widetilde{Q}_n \\
&\mathrm{s.t.} ~\mathrm{C1:}~  x_n \in\{0,1\}, x_n \left(Q_n\left(\hat{R}_n^{\rm BT}\right)-\overline{Q}_n^{\rm min}\right) \geq  0,\;\;\; \forall n, \nonumber \\
&~~~~~\mathrm{C2:}~Q_n\left(\hat{R}_n^{\rm BT}\right) \leq \gamma_n^{\rm peak}\overline{Q}_n^{\rm min},\;\;\; \forall n, \nonumber \\
&~~~~~\mathrm{C3:}~{\rm Eq. (3)},\;\; {\rm Eq. (4)},\nonumber \\
&~~~~~\mathrm{C4:}~ {\rm{C}}_{{\nu},n,i}^{\rm{BWP}} \bigcap {\rm{C}}_{{\nu'},n',i'}^{\rm{BWP}}  = \emptyset , \;\;\; \forall n, \nu, \nu'\in\{\rm BT, \rm ET\}, \nonumber \\
&~~~~~\mathrm{C5:}~\mu_{n,i}^{\nu} \in \mathbb{Z}, \eta_{n,i}^{\nu} \in \mathbb{Z}, \eta_{n,i}^{\nu} \leq 14,  \;\;\; \forall n, \nu, \in\{\rm BT, \rm ET\},\nonumber
\end{align}
where $\mathbf{x}=[x_n]$ are binary values that indicate whether UE is served or not, $\boldsymbol{\mu}=[\mu_{n,i}^{\rm BT},\mu_{n,i}^{\rm ET}]$ and $\boldsymbol{\eta}=[\eta_{n,i}^{\rm BT},\eta_{n,i}^{\rm ET}]$. Constraint $\mathrm{C1}$ makes sure the served UEs meet the minimum QoE value $\overline{Q}_n^{\rm min}$ and the solutions of problem \eqref{P1} are always feasible; $\mathrm{C2}$ shows that the QoE for entire 360$^\circ$ view at BT needs to be restricted to below a peak value $\gamma_n^{\rm peak}\overline{Q}_n^{\rm min}$ in practice, since QoE for viewport videos at ET is more important; $\mathrm{C3}$ is the limitation of available time-frequency resources; $\mathrm{C4}$ illustrates that there is no overlap among bandwidth parts; In constraint $\mathrm{C5}$, mini-slot $\eta_{n,i}^{\nu} \leq 14$ means that each slot consists of several mini-slots and there are always 14 OFDM symbols per slot regardless of the specific numerology~\cite{Zaidi2018,Ljiljana2020}.

\section{DRL-based Algorithm Design}\label{algorithm_design}
Problem \eqref{P1} is combinatorial with non-overlap constraints, which is challenging to solve.  Since deep reinforcement learning (DRL) is a powerful machine learning tool to deal with discrete decision-making problems, we provide a DRL-based solution, in which an agent (link between the UE and BS) interacts with its environment, and the key components of RL are detailed as follows:
\begin{itemize}
  \item State: In a given resource grid with time and frequency boundaries, the observed state space at BS is denoted by $\mathcal{S}=\{\boldsymbol{\Theta,\Psi,\aleph,\Phi}\}$, where $\boldsymbol{\Theta}$ contains the allocated bandwidth parts and idle ones, $\boldsymbol{\Psi}$ contains the allocated BT and ET's bandwidth parts for each UE and the corresponding QoE $Q\left(\hat{R}^{\rm BT}\right)$ and $\widetilde{Q}$ given by \eqref{eq2}, $\aleph$ contains the order of serving UE at each time-step, and $\Phi$ contains the time and frequency boundary lines of the allocated resources at the beginning of each episode. The state vector at each time-step is $\mathbf{s}_t \in \mathcal{S}$.
  \item Actions: The action $a_t$ taken by the agent include the numerology $\mu$, mini-slot $\eta$ and their corresponding bandwidth part ${\rm C}^{\rm BWP}$ given by \eqref{BWP} (as illustrated Fig. 1).  Given the state $\mathbf{s}_t$, agent chooses feasible numerology and mini-slot values based on constraints $\mathrm{C1}$ and $\mathrm{C5}$ (Note that both the objective and constraint $\mathrm{C1}$ of problem \eqref{P1} only depend on the selected mini-slot thanks to our resource block and bandwidth part designs in Section II.), and thus determines feasible bandwidth parts based on the current time and frequency boundary lines, to satisfy the constraints $\mathrm{C3}$-$\mathrm{C4}$.
  \item Reward:  When the action is executed, agent obtains a reward. Let $M$ denote the maximum allowable number of time-steps before terminating an episode, and the reward function is given by
  \begin{align}\label{reward}
  \hspace{-0.3cm}  r_{n,t}=\left\{ \begin{array}{l}
\varrho\Delta\widetilde{Q}_n\left(t\right)+ \left(1-\varrho\right)\mathcal{Z}, \; {\rm{if}}\;{\rm{C1 - C5~are~met}},\\
 {\rm \mathcal{H}},\;{\rm{if~episode~terminates}} \;{\rm{and~C1 - C5~are~met}}, \\
 \Theta,\; {\rm otherwise}, \\
 \end{array} \right.
  \end{align}
where $\Delta\widetilde{Q}_n\left(t\right)=\widetilde{Q}_{n}\left(t\right) -\widetilde{Q}_{n}\left(t-1\right) $ is the incremental value of the agent's QoE at time-step $t$, $\mathcal{Z}$ is a penalty to enable that the agent is time-aware (namely limited number of time-steps for the agent's interaction with the environment)~\cite{Boutiba2022,Fabio2018}, which is much helpful for system stability and low-latency immmersive communications, $\varrho$ is the weighting parameter,  a high reward value $ {\rm \mathcal{H}}$ is obtained if the episode terminates and constraints ${\rm{C1}}$-${\rm{C5}}$ are met, otherwise incurring a penalty $ \Theta$.
\end{itemize}
\begin{figure}[t!]
\centering
\includegraphics[width=3.2 in]{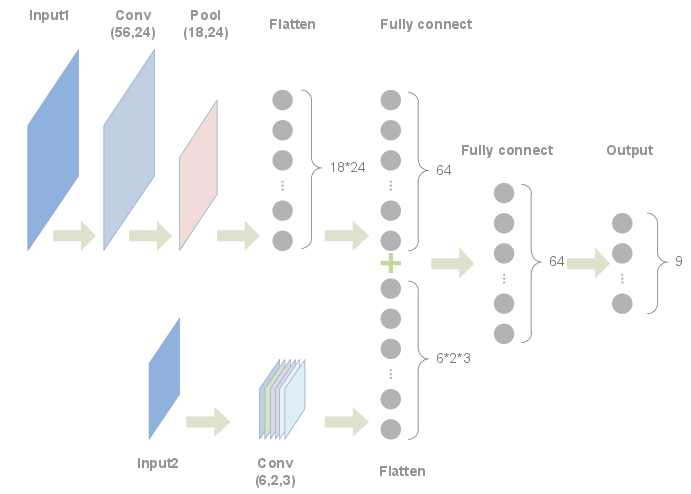}
\caption{Deep convolutional neural network structure in the proposed DQN-based method.
}% \textcolor[rgb]{1.00,0.00,0.00}{edge clouds  at or linked to the SBSs through optical fiber}
\label{Q_network}
\end{figure}

In light of deep Q-learning~\cite{Mnih2015}, we propose a DRL-based algorithm to solve the problem \eqref{P1}, which is detailed in \textbf{Algorithm 1} at next page.  The deep Q-network of \textbf{Algorithm 1} is illustrated in Fig.~\ref{Q_network}.
{\begin{algorithm}[htp] \label{algorithmic1}
  Initialize the maximum number of episodes $E$, the maximum allowable number of time-steps $M$ per episode, and deep Q-network (shown in Fig. 2);  the episode index $e=1$ \\
\While{$e \leq E$}{initialize the state space $\mathcal{S}$, in which $\aleph$ is initialized by sorting the UEs in descending order of their QoE levels under equal resource allocation;\\
\For{$t=1,\cdots,M$}{
a) Select a random action $a_{n,t}$ according to $\varepsilon$-greedy policy. Note that some actions need to be excluded when the constraints $\mathrm{C1}$-$\mathrm{C5}$ are not met;\\
b) After executing the selected action, obtain a reward given by \eqref{reward} and observe the next state $\mathbf{s}_{t+1}$;\\
\If{{\rm constraint C2 is violated}}
{the corresponding link is no longer considered as an agent at BT phase in this episode;}
c) Store the transitions $\{\mathbf{s}_t,{\mathbf{a}}_{t},\mathbf{r}_{t},\mathbf{s}_{t+1}\}$ into memory;\\
d) Train the deep Q-network  using minibatch of transitions data from the memory; \\
e) Update the policy $\pi$: $\mathcal{S} \rightarrow \mathbf{a}$;\\
}
iii) $e=e+1$;}
{The corresponding action $\mathbf{a}$ is obtained}
 \caption{{DRL based Time-frequency Resource Allocation}}
\end{algorithm} }

{\hspace{-0.5cm}\footnotesize \begin{flushleft}
\begin{table}[h]\label{table1} \footnotesize
\centering
\caption{Simulation parameters}
\setlength\tabcolsep{1.2 pt}
\hspace{-0.7cm} \begin{tabular}{|l|l|}
  \hline
  BT view coverage & $\mathcal{C}^{{\rm{BT}}} = 360^{\circ} \times 180^{\circ}$ \\ \hline
  ET view coverage & $\mathcal{C}^{{\rm{ET}}} = 135^{\circ} \times 135^{\circ}$ \\ \hline
  mmWave carrier frequency &  $f_c= 28$GHz  \\ \hline
  System bandwidth & $B=69.12$MHz  \\ \hline
  Each video frame duration & $T=0.0625$ms \\ \hline
  Effective transmit antenna gain per UE video  & $G_{\rm t}=15$dBi  \\ \hline
  Effective receive antenna gain per UE video  & $G_{\rm r}=10$dBi  \\ \hline
  Large-scale channel fading power gain &  $\left|\hbar_n\right|^2=\left( {\frac{3\times 10^8}{{4\pi f_c }}} \right)^2 \times d_n^{ - 2} $   \\ \hline
  Noise's PSD & $\delta^2=-169$dBm/Hz\\ \hline
  Total transmit PSD & $p_{\rm total}=-47$dBm/Hz \\ \hline
Vector of four UEs' minimum QoE levels &  $\overline{\mathbf{Q}}_{1\times4}=\left[4.9,4.6,4.8,4,6\right]$ \\ \hline
Peak levels of four UEs' QoE at BT &  $5\overline{\mathbf{Q}}_{1\times4}$ with $\gamma_n^{\rm peak}=5$\\ \hline
  Numerology & { $\mu=\{4,5,6\}$~\cite{Ljiljana2020} }\\ \hline
Mini-slot & { $\eta=\{2,4,7\}$~\cite{Ljiljana2020}}\\ \hline
\end{tabular}
\end{table}
\end{flushleft}
}

\section{Simulation Results}\label{sec:simulation}
This section provides numerical results to confirm the efficiency of the proposed DRL-based time-frequency resource allocation in Section III.  The learning parameters include:  learning rate is 0.001, the maximum allowable number of time-steps per episode is $M=1000$, the parameters in the reward function are set as $\varrho=0.5$, $\mathcal{Z}=-0.01$, ${\rm \mathcal{H}}=500$, and $\Theta=-2$, respectively, the discount factor $\gamma=0.99$, and deep Q-network in Fig. 2 are trained using Adam optimizer. The system parameters are detailed in Table I. {In the simulations, the communication distance $d_n\geq 1$ from the BS to the $n$-th UE is uniformly distributed with the cell radius $200$m,}  the probability of precisely predicting UE's FoV  $\rho_n$  follows the truncated normal distribution with the truncation interval [0.6,1] (its mean and variance of the parent general normal probability density function are 0.8 and 0.49, respectively), {and $a_n=0$, $b_n=1$, $\forall n$.  The results are obtained by averaging over 2000 trials.}

\subsection{Convergence}
\begin{figure}[htbp]
\centering
 \includegraphics[width=3.0 in,height=2.3 in]{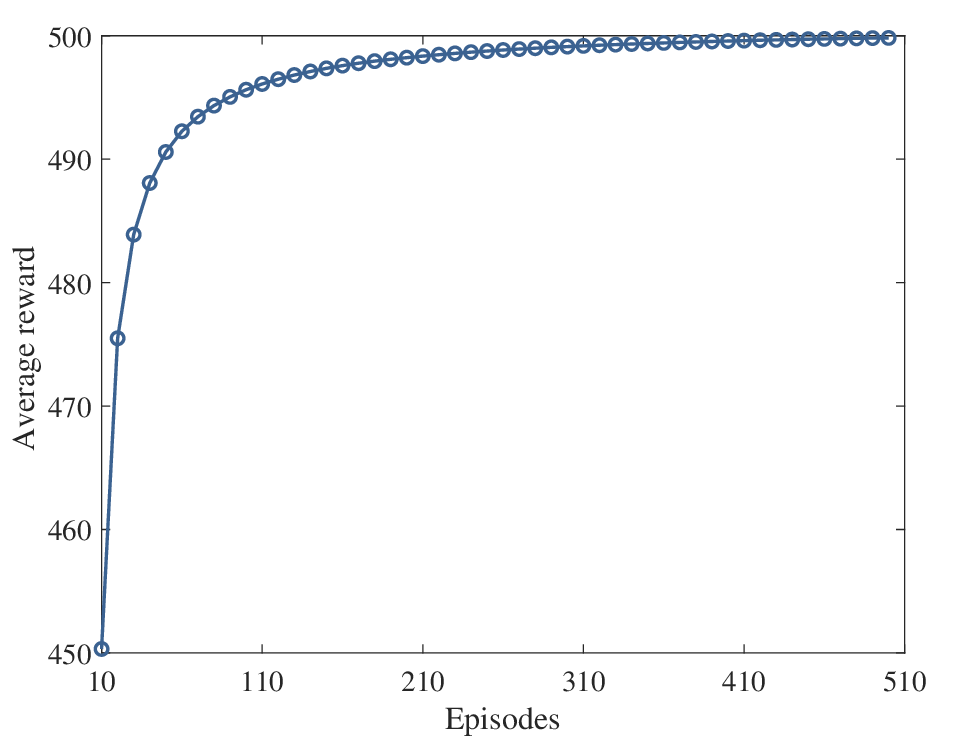}
\caption{The convergence of the proposed DRL-based method.}
 \label{fig3aa}
\end{figure}

Fig.~\ref{fig3aa} confirms that the proposed DRL-based method converges at a fast speed. Moreover, the input dimensionality is not large based on  our resource block design given by \eqref{resource_block} ($56\times24$ resource blocks in the simulations), which relieves the burden of computation and memory. The maximum average award becomes stable when the number of episodes is greater than $E=310$.
\begin{figure}[htbp]
\centering
 \includegraphics[width=3.0 in,height=2.2 in]{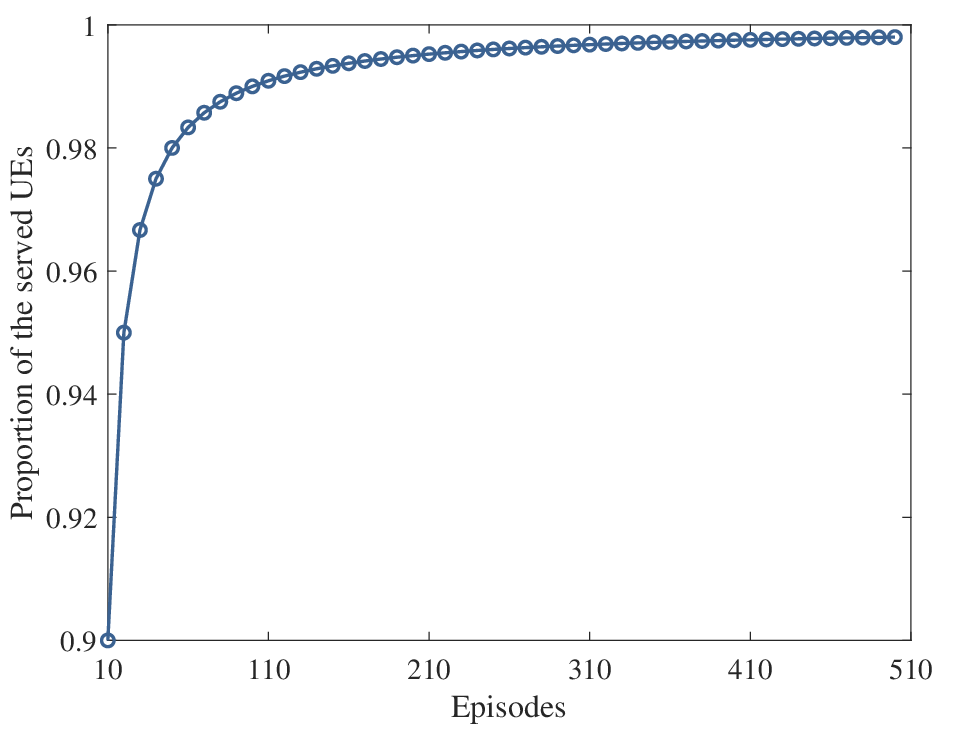}
\caption{The feasibility of the proposed DRL-based method.}
 \label{fig3bb}
\end{figure}
\subsection{Feasibility}

Fig. \ref{fig3bb} confirms that the proposed method achieves the feasibility and satisfies the constraints $\mathrm{C1}$-$\mathrm{C5}$ of problem \eqref{P1}, namely the minimum QoE of the served UEs are guaranteed, and the percentage of the served UEs is almost 100$\%$ as the number of episodes is larger than 310.

\subsection{Efficiency}
\begin{figure}[htbp]
\centering
 \includegraphics[width=3.0 in,height=2.1 in]{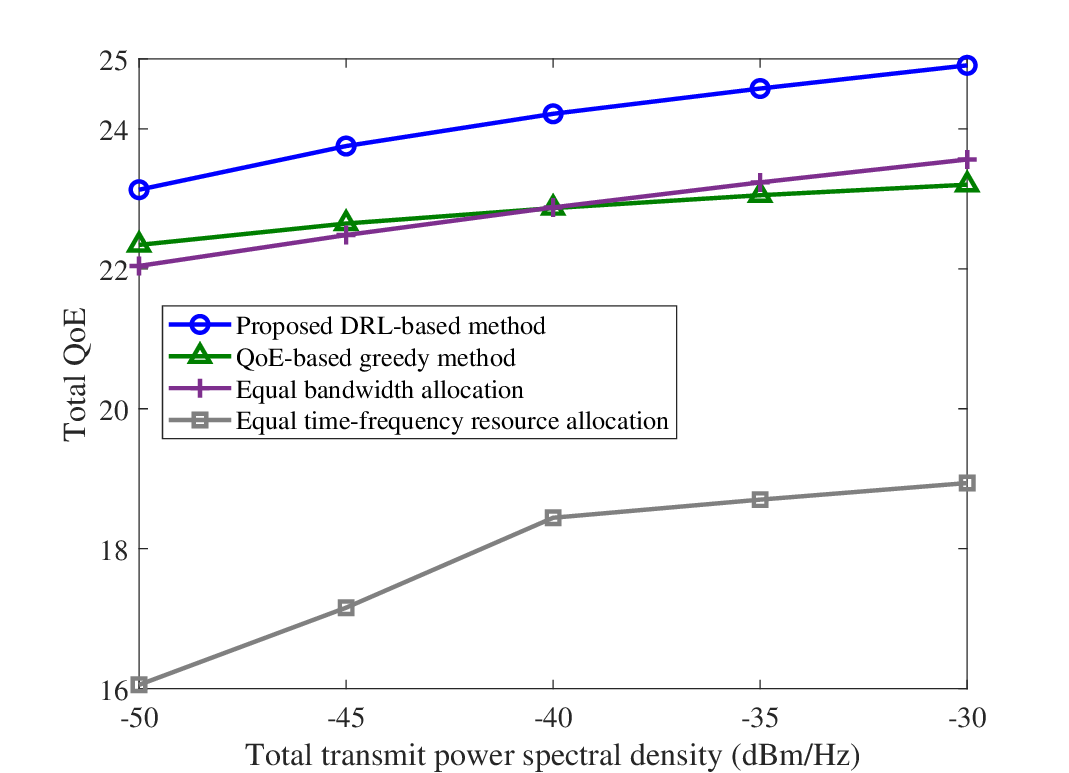}
\caption{The performance of the proposed DRL-based method.}
 \label{fig3cc}
\end{figure}
  Fig. \ref{fig3cc} confirms the efficiency of the proposed method compared to the equal resource allocation cases. In the simulations, the number of episodes is set as $E=510$ based on the empirical results in Figs. \ref{fig3aa} and \ref{fig3bb}. { There is no commonly-accepted method yet in this uncharted study and three benchmark solutions are considered as follows: 1) the QoE-based greedy method follows the approach of 3GPP, which allows UEs with higher QoE levels to take priority for resource allocation~\cite{Schwarz2010}; 2) the equal bandwidth allocation case considers free-viewpoint without FoV predication~\cite{huming2023} (BS sends the entire 360-degree videos to the UEs.); and 3) the equal time-frequency resource allocation case considers that the frequency bandwidths are equally allocated to the UEs and time of each UE's BT is the half of a video frame duration. We see that the proposed DRL-based method outperforms all the benchmark cases. The QoE-based greedy method has better performance than other benchmark cases in the low transmit PSD regime, however, it achieves lower QoE than the equal bandwidth allocation case at higher transmit PSDs since the multiuser diversity gain of the equal bandwidth allocation case is higher than the QoE-based greedy one. } In addition, the equal time-frequency resource allocation case obtains the worst QoE performance, due to the fact that in the equal time-frequency resource allocation case, the minimum QoE of some UEs cannot be met when time resources are not enough for BT transmissions, i.e., less UEs are served in this case.

\section{Conclusions}\label{conclusion_section}
In multiuser immersive communications, diverse service requirements posed a great challenge to the time-frequency resource management. To address this issue, the scalable delivery of 360-degree video was considered. By exploiting the numerology and mini-slot techniques, a flexible resource block structure has been designed, which only depends on the system's maximum and minimum numerology. A DRL-based method was developed to optimize the numerology and mini-slot values for maximizing the network QoE.
The results confirmed the advantages of our proposed method, moreover, the proposed resource block design was very useful for facilitating the performance analysis and reducing the system costs.

\bibliographystyle{IEEEtran}
%\bibliography{NEWmybib}

\begin{thebibliography}{10}
\providecommand{\url}[1]{#1}
\csname url@samestyle\endcsname
\providecommand{\newblock}{\relax}
\providecommand{\bibinfo}[2]{#2}
\providecommand{\BIBentrySTDinterwordspacing}{\spaceskip=0pt\relax}
\providecommand{\BIBentryALTinterwordstretchfactor}{4}
\providecommand{\BIBentryALTinterwordspacing}{\spaceskip=\fontdimen2\font plus
\BIBentryALTinterwordstretchfactor\fontdimen3\font minus
  \fontdimen4\font\relax}
\providecommand{\BIBforeignlanguage}[2]{{%
\expandafter\ifx\csname l@#1\endcsname\relax
\typeout{** WARNING: IEEEtran.bst: No hyphenation pattern has been}%
\typeout{** loaded for the language `#1'. Using the pattern for}%
\typeout{** the default language instead.}%
\else
\language=\csname l@#1\endcsname
\fi
#2}}
\providecommand{\BIBdecl}{\relax}
\BIBdecl

\bibitem{John_2012}
J.~G. Apostolopoulos, P.~A. Chou, B.~Culbertson, T.~Kalker, M.~D. Trott, and
  S.~Wee, ``The road to immersive communication,'' \emph{Proc. {IEEE}}, vol.
  100, no.~4, pp. 974--990, April 2012.

\bibitem{huming2023}
M.~Hu, L.~Wang, B.~Tan, and S.~Jin, ``Two-tier 360-degree video delivery
  control in multiuser immersive communications systems,'' \emph{{IEEE} Trans.
  Veh. Technol.}, vol.~72, no.~3, pp. 4119--4123, Mar. 2023.

\bibitem{Zaidi2018}
A.~A. Zaidi, R.~Baldemair, V.~Moles-Cases, N.~He, K.~Werner, and A.~Cedergren,
  ``{OFDM} numerology design for {5G} new radio to support {IoT, eMBB, and
  MBSFN},'' \emph{IEEE Commun. Stand. Mag.}, vol.~2, no.~2, pp. 78--83, 2018.

\bibitem{Peng2017}
P.~Guan, D.~Wu, T.~Tian, J.~Zhou, X.~Zhang, L.~Gu, A.~Benjebbour, M.~Iwabuchi,
  and Y.~Kishiyama, ``{5G} field trials: {OFDM}-based waveforms and mixed
  numerologies,'' \emph{{IEEE} J. Sel. Areas Commun.}, vol.~35, no.~6, pp.
  1234--1243, June 2017.

\bibitem{Lei2018}
L.~You, Q.~Liao, N.~Pappas, and D.~Yuan, ``Resource optimization with flexible
  numerology and frame structure for heterogeneous services,'' \emph{IEEE
  Commun. Lett.}, vol.~22, no.~12, pp. 2579--2582, Dec 2018.

\bibitem{xiaopei_2018}
X.~Zhang, L.~Zhang, P.~Xiao, D.~Ma, J.~Wei, and Y.~Xin, ``Mixed numerologies
  interference analysis and inter-numerology interference cancellation for
  windowed {OFDM} systems,'' \emph{{IEEE} Trans. Veh. Technol.}, vol.~67,
  no.~8, pp. 7047--7061, Aug. 2018.

\bibitem{Memisoglu2021}
E.~Memisoglu, A.~E. Duranay, and H.~Arslan, ``Numerology scheduling for {PAPR}
  reduction in mixed numerologies,'' \emph{IEEE Wireless Commun. Lett.},
  vol.~10, no.~6, pp. 1197--1201, June 2021.

\bibitem{Boutiba2022}
K.~Boutiba, M.~Bagaa, and A.~Ksentini, ``Radio resource management in
  multi-numerology {5G} new radio featuring network slicing,'' in \emph{Proc.
  IEEE ICC}, 2022, pp. 359--364.

\bibitem{Nguyen2022}
T.-S.-L. Nguyen, S.~Kallel, N.~Aitsaadi, C.~Adjih, and I.~Fajjari, ``A flexible
  numerology configuration for efficient resource allocation in {3GPP V2X 5G}
  new radio,'' in \emph{IEEE GLOBECOM}, 2022, pp. 4449--4454.

\bibitem{Hao2021}
H.~Yin, L.~Zhang, and S.~Roy, ``Multiplexing {URLLC} traffic within {eMBB}
  services in {5G NR}: {F}air scheduling,'' \emph{{IEEE} Trans. Commun.},
  vol.~69, no.~2, pp. 1080--1093, Feb. 2021.

\bibitem{Setayesh2022}
M.~Setayesh, S.~Bahrami, and V.~W. Wong, ``Resource slicing for {eMBB and
  URLLC} services in radio access network using hierarchical deep learning,''
  \emph{{IEEE} Trans. Wireless Commun.}, vol.~21, no.~11, pp. 8950--8966, Nov.
  2022.

\bibitem{Schober2022}
M.~Darabi, V.~Jamali, L.~Lampe, and R.~Schober, ``Hybrid puncturing and
  superposition scheme for joint scheduling of {URLLC and eMBB} traffic,''
  \emph{IEEE Commun. Lett.}, vol.~26, no.~5, pp. 1081--1085, May 2022.

\bibitem{Yaqoob2020}
A.~Yaqoob, T.~Bi, and G.-M. Muntean, ``A survey on adaptive 360{$^\circ$} video
  streaming: {S}olutions, challenges and opportunities,'' \emph{IEEE Commun.
  Surv. Tutor.}, vol.~22, no.~4, pp. 2801--2838, 2020.

\bibitem{Yaowang2019}
L.~Sun, F.~Duanmu, Y.~Liu, Y.~Wang, Y.~Ye, H.~Shi, and D.~Dai, ``A two-tier
  system for on-demand streaming of 360 degree video over dynamic networks,''
  \emph{IEEE J. Emerg. Sel. Topics Circuits Syst.}, vol.~9, no.~1, pp. 43--57,
  July 2019.

\bibitem{Nguyen2018}
T.~C. Nguyen and J.-H. Yun, ``Predictive tile selection for 360-degree {VR}
  video streaming in bandwidth-limited networks,'' \emph{IEEE Commun. Lett.},
  vol.~22, no.~9, pp. 1858--1861, Sept. 2018.

\bibitem{Kaigui2019}
Y.~Zhang, P.~Zhao, K.~Bian, Y.~Liu, L.~Song, and X.~Li, ``{DRL360: 360}-degree
  video streaming with deep reinforcement learning,'' in \emph{Proc. IEEE
  INFOCOM}, 2019, pp. 1252--1260.

\bibitem{Hongkai2022}
N.~Kan, J.~Zou, C.~Li, W.~Dai, and H.~Xiong, ``Rapt360: Reinforcement
  learning-based rate adaptation for 360-degree video streaming with adaptive
  prediction and tiling,'' \emph{IEEE Trans. Circuits Syst. Video Technol.},
  vol.~32, no.~3, pp. 1607--1623, Mar. 2022.

\bibitem{Chengjun2021}
C.~Guo, L.~Zhao, Y.~Cui, Z.~Liu, and D.~W.~K. Ng, ``Power-efficient wireless
  streaming of multi-quality tiled {360 VR} video in {MIMO-OFDMA} systems,''
  \emph{{IEEE} Trans. Wireless Commun.}, vol.~20, no.~8, pp. 5408--5422, Aug.
  2021.

\bibitem{Ljiljana2020}
L.~Marijanovi\'{c}, S.~Schwarz, and M.~Rupp, ``Multiplexing services in {5G}
  and beyond: {O}ptimal resource allocation based on mixed numerology and
  mini-slots,'' \emph{IEEE Access}, vol.~8, pp. 209\,537--209\,555, Nov. 2020.

\bibitem{TS38_211}
3GPP TS 38.211, ``NR; Physical channels and modulation (Release 17),'' June
  2023.

\bibitem{Fabio2018}
F.~Pardo, A.~Tavakoli, V.~Levdik, and P.~Kormushev, ``Time limits in
  reinforcement learning,'' in \emph{Pro. ICML}, 2018, pp. 1--10.

\bibitem{Mnih2015}
V. Mnih \emph{et al.}, ``Human-level control through deep reinforcement
  learning,'' \emph{Nature}, vol. 518, no. 7540, pp. 529--533, Feb. 2015.

\bibitem{Schwarz2010}
S.~Schwarz, C.~Mehlf\"{u}hrer, and M.~Rupp, ``Low complexity approximate
  maximum throughput scheduling for {LTE},'' in \emph{Asilomar Conf. Signals
  Syst. Comput.}, 2010, pp. 1563--1569.

\end{thebibliography}

\end{document}